\documentstyle{l-aa}
\newcommand{\lya}{Lyman~$\alpha$}

\newcommand{\kms}{\,km~s$^{-1}$}      
\def\ltsima{$\; \buildrel < \over \sim \;$}
\def\simlt{\lower.5ex\hbox{\ltsima}}
\def\gtsima{$\; \buildrel > \over \sim \;$}
\def\simgt{\lower.5ex\hbox{\gtsima}}
 
\begin{document}

\thesaurus{11(11.01.1; 11.09.4; 11.17.1; 11.17.4)}

\title{Zinc and Chromium Abundances in a Third Damped Lyman~$\alpha$
System at Intermediate Redshift
}

\author{Max Pettini\inst{1} and David V. Bowen\inst{2}}

\institute{
Royal Greenwich Observatory, Madingley Road, Cambridge, CB3 0EZ, UK
\and Royal Observatory, Blackford Hill, Edinburgh, EH9 3HJ, UK
      }

\maketitle
\markboth{M. Pettini \& D.V. Bowen:
Metal abundances in a galaxy at $z = 1.0093$}{}

\begin{abstract}
We have determined the metallicity of the $z_{\rm abs} = 1.0093$ damped 
\lya\ system in the bright QSO EX~0302$-$223; this is only the third such 
measurement at redshifts $z \simlt 1$\,. Unlike the previous two cases,
we find that the abundance of Zn is only a factor of $\sim 2$ lower than 
in the Galactic interstellar medium today and is entirely compatible 
with the typical metallicity of stars in the Milky Way disk at a 
look-back time of 9.5~Gyrs. Although the galaxy responsible for producing 
the absorption system has yet to be positively identified, our observations 
show that galaxies on a chemical evolution path similar to that 
of the Milky Way do contribute to the damped \lya\ population at 
intermediate redshifts. Cr is 2.5 times less abundant than Zn, presumably 
because of depletion onto dust; 
however, the 
degree of depletion is less severe than in diffuse interstellar clouds 
in the disk of our Galaxy and 
in the Magellanic Clouds. 
Evidently, the interstellar environment in damped 
\lya\ galaxies is less conducive to the formation and survival of 
dust grains (and molecular hydrogen), but the physical processes
at the root of this effect have yet to be clarified.
\vspace{-0.1cm}
\end{abstract}

\section{Introduction}
Damped \lya\ systems (DLAs) are commonly considered to be
the high redshift progenitors of present-day luminous galaxies 
(e.g. Wolfe 1990). 
If this is the case, 
they present us with some of the best opportunities for studying in detail 
the physical conditions of galaxies at early stages of evolution, 
including kinematics, chemical abundances, dust content, temperature, 
and densities of both particles and radiation
(e.g. Prochaska \& Wolfe 1997; Lu et al. 1996; Pettini et al. 1997a,b; 
Ge, Bechtold, \& Black 1996; Ge \& Bechtold 1997).
In the past few years there has been much emphasis in particular 
on measuring element abundances in the interstellar gas of DLA absorbers with 
the long term aim of tracing the chemical evolution of the universe 
from redshift $z \approx 4$ to the present time.
Zinc is an especially useful tracer of metallicity
because:~ {\it (a)} it shows little affinity for interstellar 
dust and is found 
in the gas-phase of the Milky Way ISM in near-solar proportions, 
and~ {\it (b)} it tracks closely the abundance of iron in stars of all 
metallicities (Pettini et al. 1990, 1997a and references therein).
The large survey  
recently completed by Pettini et al. (1997b)
has shown that at redshifts $z \simeq 1.5-3$ the typical 
abundance of Zn in DLAs is 
${\rm [\langle Zn/H \rangle]_{DLA}}  \simeq -1.1$, 
or approximately 1/13 of solar\footnote{In this paper we use
the conventional notation 
where
[X/Y]~=log~(X/Y)$-$log~(X/Y)$_{\sun}$}, reflecting 
a generally low degree of chemical enrichment in the universe at 
look-back times of $\approx 11 - 14$~Gyr 
($H_0 = 50$~\kms\  Mpc$^{-1}$ and  $q_0  =  0.01$ adopted throughout
the paper).
There are indications that 
${\rm [\langle Zn/H \rangle]_{DLA}}$ is
even lower at higher redshifts; this 
led Pettini et al. (1995) to propose that  
$z \simeq 3$ may have been the period when the 
first major episodes of star formation 
took place in galaxies.
Such a  conclusion 
is supported
by the results of deep imaging surveys in the rest frame ultraviolet 
which also indicate a marked increase in the comoving star 
formation rate from $z \simeq 4$ to  $\simeq 2.75$ 
(Steidel et al. 1996; Madau et al. 1996).

In this picture the typical metallicity of DLAs should rise to near-solar  
values at redshifts $z \simlt 1$, when the cosmic star formation rate 
was near its peak (Lilly et al. 1995; Madau 1996) and 
when stars forming in the disk of our Galaxy had 
[Fe/H] $\simgt -0.5$ 
(Edvardsson et al. 1993; Friel 1995).
Verifying whether this is indeed the case is a crucial
element in making the connection between DLA absorbers and the 
population of normal galaxies; 
yet this seemingly simple test has proved difficult
up to now.
The main problem is the paucity of known DLAs at 
intermediate and low redshifts (e.g. Bahcall et al. 1996), 
due to the combination of three effects: the reduced redshift path, the 
general consumption of H~I gas into stars in the universe 
(Lanzetta, Wolfe \& Turnshek 1995), 
and the need for space observations to identify 
a damped \lya\ line at $z \simlt 1.5$\,.

To date measurements of metallicity have been published for only two 
damped systems at $z < 1$: at $z_{\rm abs} = 0.6922$ 
in Q1328$+$307 (3C~286---Meyer \& York 1992) and at
$z_{\rm abs} = 0.8596$ in Q0454$+$039
(Steidel et al. 1995).
The fields of both QSOs have been imaged from the ground 
(Steidel et al. 1994, 1995) and with the {\it Hubble Space Telescope}
(Le Brun et al. 1997), and plausible candidates for the absorbing galaxies 
have been identified. 
However, in neither case is a Milky Way-type galaxy
indicated.
The former appears to be an extended objects of low surface brightness
and low metallicity ([Zn/H] $= -1.21$), 
consistent with the reduced  
star formation efficiencies typical of such galaxies (McGaugh 1994).
The latter also has a relatively low abundance, 
[Zn/H] $= -0.85$,
and a rest frame $B$-band
absolute magnitude, $M_B \simeq -19.5$, corresponding to 
$\approx 0.25 L^{\ast}$.
The finding that the only two abundance 
measurements available at $z < 1$ are both well below 
solar---and quite typical of the sample at $z > 1.5$, 
has caused concerns as to whether  
DLAs are really unbiased tracers of the 
chemical evolution history of the universe.

Here we report measurements of the abundances of Zn and Cr 
in a third DLA at intermediate redshift, 
the $z_{\rm abs} = 1.0093$ system in the QSO EX~0302$-$223.
This absorber turns out to have a metallicity more in line 
with that of the Milky Way at a look-back time of 9.5~Gyr. 
While it is reassuring to confirm that galaxies with a chemical enrichment 
similar to that of the Galactic disk 
do indeed contribute to the cross-section for damped \lya\ absorption at 
these redshifts, the new data also stress the 
importance of building a reasonably large sample of such measurements 
before drawing conclusions on the nature of the DLA population.

\section{The $z_{\rm abs} = 1.0093$ DLA Absorber in EX~0302$-$223} 

The serendipitous {\it Einstein} X-ray source EX~0302$-$223
was identified by Chanan, Margon, \& Downes (1981) with an 
optically bright 
($V = 16.0$) $z_{\rm em} = 1.400$ QSO.
Lanzetta et al. (1995) drew attention to three possible 
damped \lya\ lines in low resolution {\it IUE} spectra
of the QSO, at $z_{\rm abs} = 0.9690$, 0.9874 and 1.0140;
an {\it HST} spectrum, which we have retrieved from the Hubble Data Archive, 
confirms that the third system is indeed damped.
EX~0302$-$223 was observed with {\it HST}  
on 1995 December 7; using the G270H grating and red
detector of the Faint Object Spectrograph (FOS), a 2390~s exposure was
recorded 
through the 0.43 arcsec circular aperture. 
The spectrum covers the wavelength interval 2225--3275~\AA;
a portion centred on the damped \lya\ absorption line
is shown in the top panel of Figure 1.
In producing this spectrum, we resampled the pipeline
calibrated data to the original dispersion of
0.51~\AA~pix$^{-1}$, 
and found it necessary to apply 
a small correction for scattered light
($\sim 2$\% of the continuum)
to bring the core of the absorption line to zero flux.

The profile of the damped \lya\ line is well fitted by a column density of 
neutral hydrogen
$N$(H$^0$)$~= {\rm (}2.15 \pm 0.35{\rm )} \times 10^{20}$~cm$^{-2}$
at $z_{\rm abs} = 1.0099$\,.
The difference between this redshift and 
$z_{\rm abs} = 1.0093$, 
measured by Petitjean \& Bergeron (1990)
from the centroids of the Mg~II and Fe~II absorption lines,
corresponds to approximately one third of a diode on the FOS detector and is 
typical of the accuracy with which the zero point of 
the FOS wavelength scale can be determined.

\begin{figure*}  
\input epsf
\centering 
\leavevmode 
\epsfxsize=1.9\columnwidth
\epsfysize=1.2\columnwidth 
\epsfbox[65 226 504 553]{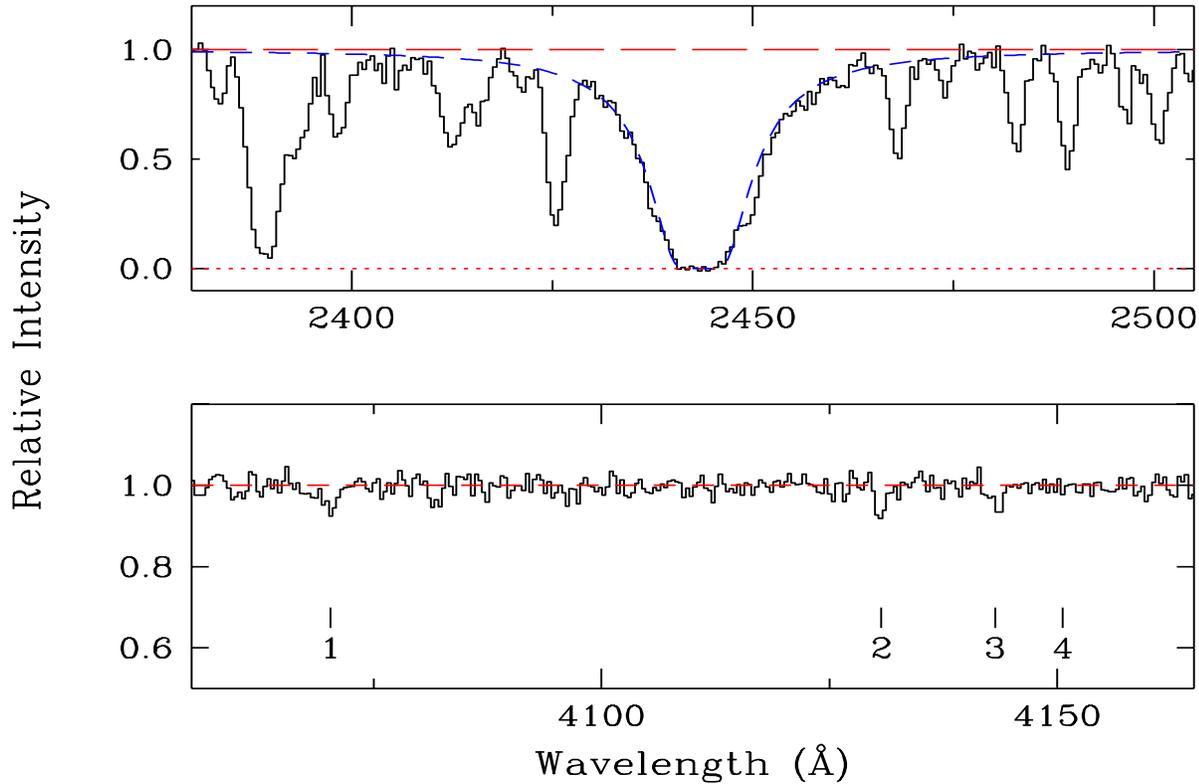}
\caption{
{\it Upper Panel:\/} Portion of the G270H FOS spectrum of 
EX~0302$-$223 encompassing the 
damped \lya\ line in the $z_{\rm abs} = 1.0093$ system. 
The resolution is 1.8~\AA\ FWHM and S/N $\simeq  20$.
The short-dash line shows the theoretical damping profile
for a column density
$N$(H$^0$)$~= 2.15 \times 10^{20}$~cm$^{-2}$\,.
~{\it Lower Panel:\/} Portion of the WHT  
spectrum covering the region of 
the Zn~II and Cr~II absorption lines in the 
DLA system.
The resolution is 0.88~\AA\ and S/N = 50.
The vertical tick marks indicate
the expected positions of the lines, as follows (vacuum wavelengths):
1:~Zn~II~$\lambda 2026.136$; 2:~Cr~II~$\lambda 2056.254$; 
3:~Cr~II~$\lambda   2062.234$~+~Zn~II~$\lambda  2062.664$  (blended);  
and 4:~Cr~II~$\lambda 2066.161$.  
Both spectra have been normalised to the
underlying continua; 
the spectrum in the lower panel is shown on an expanded vertical scale.  
}
\end{figure*}

Le Brun et al. (1997) have obtained images of the QSO field
through the F702W (2400~s integration) 
and F450W (2000~s integration) filters of WFPC2 on the {\it HST}.
After subtracting the instrumental point spread function, 
they suggest that the two galaxies closest to the QSO sight-line 
(objects 2 and 3 in their Figure 3, respectively
1.1 and 2.5 arcsec from the QSO)
are promising candidates for the damped absorber.
However, a positive identification will have to wait until spectra, or at 
least broad-band colours, are available. 

If the two galaxies are indeed at $z \simeq 1.0093$,
the corresponding impact parameters are  
$12~h_{50}^{-1}$ and $27~h_{50}^{-1}$~kpc
respectively. 
Le Brun et al. measured magnitudes 
$m_{702} = 25.4$ and 23.8 (on a Vega-based photometric system); 
at $z = 1.0093$ these values would correspond to 
$M_B = -19.3$ and $-20.9$
for galaxy 2 and 3 respectively, adopting a k-correction appropriate to
an Sbc spectral template 
(we used the template by Coleman et al. 1980 for 
consistency with Le Brun et al.). 
Thus, if at $z = 1.0093$, 
galaxy 2  has a luminosity $L_B  \approx 0.2 L_B^{\ast}$,
while object 3 is more luminous
with $L_B  \approx L_B^{\ast}$.
Both galaxies appear compact with 
approximate dimensions $4 \times 3 $
and $1 \times 2$~kpc respectively; 
presumably we are seeing predominantly their bulges 
at the exposure levels attained 
in the WFPC2 images.
If {\it both} galaxies are at $z = 1.0093$ their projected 
separation from each other is only $22~h_{50}^{-1}$~kpc,
and they may well be interacting. 
Thus the galaxies may be intrinsically fainter 
than their luminosities suggest, 
if we are observing them during a period of enhanced star formation 
triggered by the interaction, and the absorbing gas in front of the QSO 
may be tidally disrupted material.
There is clearly a strong incentive to determine the redshifts of 
galaxies 2 and 3 and clarify the nature of this damped
Lyman~$\alpha$ system.

At least one object in the field of EX~0302$-$223
is at a redshift close to that of the DLA. Le Brun et al. refer to 
unpublished observations of object 7 by Guillemin \& Bergeron
showing that it is at redshift $z = 1.000$; at a separation of
$84~h_{50}^{-1}$~kpc from the QSO sight-line, however, 
this galaxy is unlikely to be the 
absorber.

\section{Zn and Cr Abundances}

We observed EX~0302$-$223 in service mode during the 
nights of 8--9 and 17--18 January 1997
with the double spectrograph at the cassegrain 
focus of the William Herschel telescope on the island of la Palma, 
Canary Islands. Table 1 gives the journal of observations. 

   \begin{table}
      \caption{Journal of Observations}
         \label{Table1}
      \[
         \begin{array}{cccc}
            \hline
            \noalign{\smallskip}
 {\rm ~~UT~Date~~} & {\rm ~~Exp.~Time~~(s)~~} 
 & {\rm ~~Resolution~~(\AA)~~} & {\rm ~~S/N~~} \\
            \noalign{\smallskip}
            \hline
            \noalign{\smallskip}
            1997~{\rm Jan}~08 & 2000 & $0.88$  & 33  \\
            1997~{\rm Jan}~09 & 2000 & $0.88$  & 28  \\
            1997~{\rm Jan}~17 & 2000 & $0.90$  & 16  \\        
            1997~{\rm Jan}~17 & 2000 & $0.90$  & 14  \\
            \noalign{\smallskip}
            \hline
         \end{array}
      \]
   \end{table}

A detailed description of the procedures used in acquiring and reducing 
the data can be found in Pettini et al. (1994). Briefly, 
we used a 1200 grooves~mm$^{-1}$ grating, blazed near 4100~\AA, 
and a thinned $1024 \times 1024$ pixel Tektronix CCD to record a
$\approx400$~\AA\ wide portion of the spectrum centred at 4121~\AA\ 
and encompassing the wavelengths of the
Zn~II~$\lambda\lambda 2026.136, 2062.664$ and
Cr~II~$\lambda\lambda 2056.254, 2062.234, 2066.161$ multiplets
in the damped \lya\ system.
A total of four exposures, each 2000~s long, was secured. 
The QSO was moved by a few arcseconds along the slit between each 
exposure, so as to use different rows of the detector to record the 
spectrum. 
Using standard IRAF routines, 
the individual spectra were optimally extracted and 
wavelength 
calibrated by reference to the emission lines of Cu-Ar and Cu-Ne hollow 
cathode lamps; they were then mapped onto a common, 
vacuum heliocentric wavelength scale before being co-added.
The QSO was observed at large zenith distances (ZD $\simeq 55 - 65$ 
degrees); the bright of moon and poor seeing 
on 17 January account for the lower S/N of those data (see Table 1).
The final spectrum has S/N = 50 and a resolution of 0.88~\AA, sampled 
with 2.2 wavelength bins.
The corresponding $3 \sigma$ limit for the rest-frame equivalent width of 
an unresolved absorption line is $W_0$($3\sigma$) = 26~m\AA.
The lower panel of Figure 1 shows the wavelength region of interest after 
division by the underlying QSO continuum; the wavelengths and rest-frame 
equivalent widths of the Zn~II and Cr~II lines are listed in Table 2.
We detect features 1, 2 and 3 
at the $\sim 5 \sigma$ level, while feature 4 
is below our detection limit.
The weakness of the absorption lines and their equivalent width ratios indicate 
that saturation effects are probably not important.

   \begin{table}
      \caption{Zn~II and Cr~II Absorption Lines}
         \label{Table2}
      \[
         \begin{array}{clllll}
            \hline
            \noalign{\smallskip}
 {\rm ~~Line~~} & ~~{\rm Ion}~~ & ~~\lambda_0^{a,b}~{\rm (\AA)}~~ 
 & ~~\lambda_{\rm obs}^{a,c}~{\rm (\AA)}~~& ~~z_{\rm abs}^a~~ 
 & ~~W_0^d~{\rm (m\AA)}~~\\
            \noalign{\smallskip}
            \hline
            \noalign{\smallskip}
            1 & {\rm Zn~II}~  & ~2026.136  & ~4071.09  & 1.0093  & ~~53 \pm 10 \\
            2 & {\rm Cr~II}~  & ~2056.254  & ~4131.55  & 1.0093  & ~~51 \pm 10 \\
            3 & {\rm Cr~II+}~ & ~2062.234+ & ~4144.01  & {\rm blend} & ~~48 \pm 10  \\        
              & {\rm Zn~II}~  & ~2062.664  &          &          &            \\
            4 & {\rm Cr~II}~  & ~2066.161  &  ~~\ldots   & ~~\ldots  & ~~\leq 30  \\
            \noalign{\smallskip}
            \hline
         \end{array}
      \]
\begin{list}{}{}
\item[$^{\rm a}$] All wavelengths are in vacuum
\item[$^{\rm b}$] Rest wavelength 
\item[$^{\rm c}$] Observed wavelength (heliocentric)
\item[$^{\rm d}$] Rest frame equivalent width 
\end{list}
   \end{table}

   \begin{table*}
      \caption{Column Densities and Abundances in the $z_{\rm abs} = 1.0093$ DLA System}
         \label{Table3}
            \[
         \begin{array}{cccccccc}
            \hline
            \noalign{\smallskip}
 ~~N{\rm (H}^0{\rm )}~~~~ & 
 ~~N{\rm (Zn}^+{\rm )}~~ &
 ~~N{\rm (Zn}^+{\rm )}/N{\rm (H}^0{\rm )}~~ &
 ~~{\rm [Zn/H]}~~~~ & 
 ~~N{\rm (Cr}^+{\rm )}~~ & 
 ~~N{\rm (Cr}^+{\rm )}/N{\rm (H}^0{\rm )}~~ &
 ~~{\rm [Cr/H]}~~~~ &
 ~~{\rm [Cr/Zn]}~~  \\
            \noalign{\smallskip}
 {\rm (}10^{20}~{\rm cm^{-2})}~~ & 
 {\rm (}10^{12}~{\rm cm^{-2})} &
 {\rm (}10^{-8}~{\rm cm^{-2})} &
 & 
 {\rm (}10^{12}~{\rm cm^{-2})} & 
 {\rm (}10^{-8}~{\rm cm^{-2})} &
 &
\\
            \noalign{\smallskip}
            \hline
            \noalign{\smallskip}
            2.15 \pm 0.35~~~~  & 
            3.0 \pm 0.6  & 
            1.4 \pm 0.4  & 
            -0.51^{+0.11}_{-0.14}~~~~ & 
            13.0 \pm 2.5 & 
            6.0 \pm 1.5  &
            -0.90^{+0.09}_{-0.13}~~~~ &
            -0.39^{+0.11}_{-0.14} \\
            \noalign{\smallskip}
            \hline
         \end{array}
      \]
   \end{table*}

Assuming that Zn~II~$\lambda 2026.136$ and 
Cr~II~$\lambda 2056.254$ are on the linear part of the curve of 
growth and adopting the $f$-values measured by Bergeson \& Lawler (1993), 
we deduce the values of column density of Zn$^+$ and Cr$^+$
listed in columns 2 and 5 
of Table 3 respectively. 
Dividing by $N$(H$^0$) and comparing with the solar abundances 
of Zn and Cr 
(Anders \& Grevesse 1989)\footnote{log~(Zn/H)$_{\sun} = -7.35$, 
log~(Cr/H)$_{\sun} = -6.32$} we reach the conclusion that
Zn and Cr are 
less abundant in the $z_{\rm abs} = 1.0093$
DLA than in the Sun by factors of
3 and 8 respectively.

\section{Discussion}

\subsection{Chemical Evolution}

The Zn abundance deduced is 
in good agreement with expectations for  
a galaxy like the Milky Way at this redshift.
In the present day interstellar medium 
[Zn/H]$_{\rm gas}= -0.19$ which may imply that
$\approx 35\%$ of Zn is locked up in dust grains
(Roth \& Blades 1995; Sembach et al. 1995).
The finding that [Zn/H]$_{\rm DLA}$ is only a factor of $\sim 2$ lower
than this value at a lookback time of 9.5~Gyrs
is consistent with the mild increase with time
of the average [Fe/H] of Galactic 
disk stars over this period (see Figure 14 of Edvardsson et al. 1993).
While it would be of great interest to relate the metallicity 
we have determined to other 
properties of the galaxy producing the DLA---such as mass, morphology and impact 
parameter---it is difficult to proceed further pending a 
positive identification of the absorber. 
Nevertheless, the observations presented here do demonstrate that galaxies 
which are on a chemical evolution path similar to that of the Milky Way 
disk do contribute to the DLA population at intermediate 
redshifts. 

The three measurements of [Zn/H]$_{\rm DLA}$ 
at $z \simlt1$ available at present
span a factor of $\sim 5$, 
from 1/3 solar found here to 1/16 solar at  
$z_{\rm abs} = 0.6922$ in 3C~286 (Meyer \& York 1992, corrected for the 
more recent $f$-values used in the present analysis).
Such a range in the degree of chemical enrichment 
is in line with indications from the imaging survey by  
Le Brun et al. (1997) that DLAs at these redshifts
arise in a diverse population of galaxies, 
which includes amorphous low-surface brightness objects and compact 
galaxies, as well as apparently normal spirals of various sizes and 
luminosities. The same presumably applies at
$z \simgt 1.5$ where values of [Zn/H]$_{\rm DLA}$ span more than one order 
of magnitude (Pettini et al. 1997b).

The challenge for the future, then, is to assess whether there are 
significant changes with redshift in the relative contributions of 
different galaxies to the cross-section for DLA absorption.
It is possible, for example, that typical spiral galaxies like the Milky 
Way become progressively under-represented with decreasing redshift,
if much of the gas has been turned 
into stars by $z \simeq 1$, or if the build-up of dust which goes hand in 
hand with production of metals introduces significant selection 
effects (e.g. Pei \& Fall 1995).
Given the wide range of values of Zn abundance encountered at all redshifts,
it is clear that tracing the global 
chemical evolution of the DLA population from high redshifts to 
$z < 1.5$  will only be possible with a moderately large 
sample of data.
The availability of STIS on the {\it HST} has now brought within reach  
Zn and Cr absorption lines at redshifts $z \simlt 0.65$; this new
opportunity should double the number of measurements in the 
next two years. Looking further ahead, the 2dF and Sloan surveys 
will lead to major advances by 
increasing by one order of magnitude existing samples of damped \lya\ 
systems.

\subsection{Dust Depletion of Chromium}

As can be seen from the last column of Table 3, 
Cr is less abundant than Zn by a factor of 
$\sim2.5$; as discussed by Pettini et al. (1997a)
the most straightforward interpretation of this abundance difference is 
the selective depletion of  Cr---and presumably other refractory 
elements---onto dust grains.
The value [Cr/Zn] = $-0.39$ when [Zn/H] = $-0.51$ fits in well 
with the broad trend of increasing depletion with increasing metallicity 
found in the analysis of 18 such measurements
by Pettini et al. (1997a---see their Figure 1).
In agreement with that study, we find that the depletion of Cr in the 
$z_{\rm abs} = 1.0093$ damped system is less severe than 
in diffuse interstellar clouds in the Galactic disk, where typically
[Cr/Zn] $\simlt -1$, 
and closer to values encountered in halo clouds located at distances of 
more than 300 pc from the plane (Savage \& Sembach 1996).

This could be simply an indication that the line of sight to 
EX~0302$-$223 does not intersect interstellar clouds of sufficiently high
density to maintain a large fraction of Cr in solid form, as is the case in the
disk of the Milky Way. Indeed, we do not know whether the 
$z_{\rm abs} = 1.0093$ DLA arises in a galactic disk at all.
On the other hand, it is intriguing that {\it none} of the 19 cases 
in which [Cr/Zn] has now been measured exhibit the degree of Cr depletion
typical of local disk clouds. 
Even along sightlines to stars in the Magellanic Clouds, 
which probe long pathlengths through both the disk and halo of the Milky Way 
in a geometry presumed to be similar to that of damped 
\lya\ systems, it is found that [Cr/Zn] $\simlt -1$, 
and there is essentially no 
overlap with the values measured in DLAs (Roth \& Blades 1997).

The lower depletions of refractory elements 
in halo clouds are commonly interpreted as an 
indication that grain processing in interstellar shocks is either more 
efficient or more frequent in the low density regions away from the 
Galactic plane  
(e.g. Savage  \& Sembach 1996).
Possibly the interstellar medium of DLA galaxies is a more hostile 
environment to interstellar grains, 
particularly if the galaxies are forming stars at 
a higher rate than the Milky Way today and consequently experience 
more frequent supernova explosions. 
Ge \& Bechtold (1997) used the 
relative populations of the rotationally excited levels of molecular 
hydrogen to measure the interstellar radiation field density near 1000~\AA\
in the $z_{\rm abs} = 1.9730$ DLA system in Q0013$-$004 and indeed found 
it to be a few times higher 
than the ambient value in the solar vicinity. 
Even so, given that the DLA population probably encompasses a wide range 
of galaxy types at different evolutionary stages,
it is implausible that all galaxies are observed during periods of 
enhanced star formation. 

Possibly several factors are at play. Clarifying the physical reasons 
at the root of the lower depletions of refractory elements 
in damped \lya\ systems---and the  
undoubtedly related lower concentrations of 
molecular hydrogen---remains an important goal for our 
understanding of the interstellar medium of distant galaxies.

\begin{acknowledgements}
We are very grateful to John Telting and Chris Benn who carried out the 
service observations on which this paper is based, and to Jim Lewis for 
his assistance in the  
reduction of the optical spectra. Mark Dickinson
kindly helped with the estimation of absolute magnitudes from {\it HST}
photometry.
\end{acknowledgements}


\begin{thebibliography}{} 
\bibitem{} Anders, E., \& Grevesse, N. 1989, Geochim. Cosmochim. Acta, 53, 197
\bibitem{} Bahcall, J.N., et al. 1996, ApJ, 457, 19
\bibitem{} Bergeson, S.D., \& Lawler, J.E. 1993, ApJ, 408, 382
\bibitem{} Chanan, G.A., Margon, B., \& Downes, R.A. 1981, ApJ, 243, L5
\bibitem{} Coleman, G.D., Wu, C-C., \& Weedman, D.W. 1980, ApJS, 43, 393
\bibitem{} Edvardsson, B., Andersen, J., Gustafsson, B., Lambert, D.L., Nissen, 
P.E., \& Tomkin, J. 1993,  A\&A, 275, 101
\bibitem{} Friel,  E.D. 1995, ARAA, 33, 381
\bibitem{} Ge, J., \& Bechtold, J. 1997, ApJ, 477, L73
\bibitem{} Ge, J.,  Bechtold, J., \& Black, J.H. 1996, ApJ, 474, 67 
\bibitem{} Lanzetta, K.M., Wolfe, A.M., \& Turnshek, D.A. 1995, ApJ, 440, 435
\bibitem{} Le Brun, V. Bergeron, J., Boisse, P., \& Deharveng, J.M. 
1997, A\&A, in press
\bibitem{} Lilly, S.J., Tresse. L., Hammer, F., Crampton, D., 
\& Le F\`{e}vre, O. 1995, ApJ, 455, 108
\bibitem{} Lu, L., Sargent, W.L.W., Barlow, T.A., Churchill, C.W., \& 
Vogt, S.S. 1996, ApJS, 107, 475
\bibitem{} Madau, P., Ferguson, H.C., Dickinson, M., Giavalisco, M., 
Steidel, C.C., \& Fruchter, A. 1996, MNRAS, 283, 1388
\bibitem{} McGaugh, S.S. 1994, ApJ, 426, 135
\bibitem{} Meyer, D.M., \& York, D.G. 1992, ApJ, 399, L121
\bibitem{} Pei, Y.C., \& Fall, S.M. 1995, ApJ, 454, 69
\bibitem{} Petitjean, P., \& Bergeron, J. 1990, A\&A, 231, 309
\bibitem{} Pettini, M., Boksenberg, A., \& Hunstead, R.W. 1990,
ApJ, 348, 48
\bibitem{} Pettini, M., King, D.L., Smith, L.J., \& Hunstead, R.W. 1995, 
in QSO Absorption Lines, ed. G. Meylan (Berlin: Springer-Verlag), 71 
\bibitem{} Pettini, M., King, D.L., Smith, L.J., \& Hunstead, R.W. 1997a,
ApJ, 478, 536
\bibitem{} Pettini, M., Smith, L.J., Hunstead, R.W., \& King, D.L. 1994,
ApJ, 426, 79
\bibitem{} Pettini, M., Smith, L.J., King, D.L. \& Hunstead, R.W. 1997b,
ApJ, 486, in press
\bibitem{} Prochaska, J.X., \& Wolfe, A.M. 1997, ApJ, in press
\bibitem{} Roth, K.C., \& Blades, J.C. 1995, ApJ, 445, L95
\bibitem{} Roth, K.C., \& Blades, J.C. 1997, ApJ, 474, L95
\bibitem{} Savage, B.D., \& Sembach, K.R. ARAA, 34, 279
\bibitem{} Sembach, K.R., Steidel, C.C., Macke, R.J., \& Meyer, D.M. 
1995, ApJ, 445, L27
\bibitem{} Steidel, C.C.,  Pettini, M., Dickinson, M., \& Persson, S.E. 
1994, AJ, 108, 2046
\bibitem{} Steidel, C.C., Bowen, D.V., Blades, J.C., \& Dickinson, M. 1995, 
ApJ, 440, L45
\bibitem{} Wolfe, A.M. 1990, in  The Interstellar Medium in Galaxies,  
ed. H.A. Thronson,  \& J.M. Shull (Dordrecht: Kluwer Academic Publishers), 387     
\end{thebibliography}
\end{document}